\documentstyle[12pt,epsfig]{article}
\newcommand{\vect}[1]{\mbox{\boldmath$#1$}}
\begin{document}
\begin{center}
{\large\bf Band Crossing studied by GCM with 3D-CHFB}\\
\vspace{0.1cm}
{\large Takatoshi Horibata}$^{\rm a \, c}$,
{\large Makito Oi}$^{\rm b \, c}$, {\large Naoki Onishi}$^{\rm b \, c}$
\\
\vspace{0.1cm}
$^{\rm a}$ 
{\it Department of Information System Engineering, 
Aomori University,
\\
Aomori-city, Aomori 030-0943, Japan}
\\
$^{\rm b}$ 
{\it Institute of Physics, College of Arts and Sciences, University of Tokyo
\\
Komaba, Meguro-ku, Tokyo 153-0041, Japan}
\\
$^{\rm c}$ 
{\it Cyclotron Laboratory, Institute of Physical and Chemical Research (RIKEN)
\\
Hirosawa 2-1, Wako-city, Saitama 351-0198, Japan}
\\
\end{center}

In $^{180}$W and $^{182}$Os nuclei, it is pointed out
[\ref{Wa93}] that three levels, g-, s- and high-$K^{+}$($8^{+}$)
bands may cross in a certain angular momentum region ($\sim 14 \hbar$).
A similar phenomenon is recently observed in $^{184}$Os nucleus
involving $K=10^{+}$ band[\ref{Sh97}].
The spectrum exhibiting signature inversion is implied to be caused
by a rather strong inter-band interaction between such a high-$K$ band
and s-band.
With the intention of understanding the inter-band interaction,
we have been studying the wobbling motion in terms of self-consistent
three-dimensional cranked Hartree-Fock-Bogoliubov (3D-CHFB) calculation
[\ref{HO94}],
together with the generator coordinate method (GCM)[\ref{HOO95}].
In the course of study, we found a stationary state of tilted axis
rotation (TAR) in the 3D-CHFB solutions.

In the previous calculation, we used a parameter set in which the
pairing interaction for neutron and the quadrupole
interaction were rather strong. 
Consequently the pairing of proton was destroyed
before the neutron at the principal axis rotation (PAR).
Although rotation alignment (RA) of high-$j$ proton orbital was once
speculated to give rise to the back-bending[\ref{NLM76}],
the possibility was examined by successive experiments that
it is due to neutron orbital.

In this paper, we present results of an improved calculation
of GCM based on the 3D-CHFB in the following two points.
1) We reduced the strength of neutron pairing interaction so as to
make RA of neutron occur in PAR, which is considered to be yrast state.
2) We use the constrained GCM to obtain solutions in a more stable way.
In a similar manner as the previous papers, we obtained PAR solutions
by cranking up along $x$-axis perpendicular to the symmetry axis
of the non-rotating state and also by cranking down along the axis.
The two bands cross at $J \sim 20 \hbar$ in the present case.

We tilted rotating axis along the prime-meridian, which intersects with
$z$- and $x$-axis on the sphere of $J=18 \hbar$.
As well as our previous calculation, the existence of the stable 
TAR state is confirmed.
The TAR minimum takes place at $\theta=24^\circ$, and its depth is
$V_{\rm TAR}=-0.166$MeV which is less than half of our previous value.
In the TAR state, proton pairing is broken (s$_{\rm p}$-band) like the 
previous calculation,
and hence level crossing occurs at least twice between the
north-pole (or south-pole) and the PAR  in the equatorial plane. 
We found the potential energy near the north-pole smaller than
the value of PAR state. 

In order to study the inter-band interaction, the wobbling motion
is investigated in terms of GCM taking the north and the south latitudes
as a generator coordinate.
Following the prescription described in ref.[\ref{BFH90}] in  solving the 
GCM equation, the expectation values of
the scalar of squared angular momentum vector $\vect{J}^{2}$ and the 
particle numbers are constrained.
We solved the following constrained Hill-Wheeler equation,
\begin{equation}
\sum_{k^{\prime}} \left[ H_{k,k^{\prime}}-\lambda_{\tau}^{(\alpha)}N^{\tau}_{k,k^{\prime}}
             -{\mu}^{(\alpha)} \vect{J}^{2}_{k,k^{\prime}}\right]
        g_{k^{\prime}}^{(\alpha)}  =     
       E^{(\alpha)}g_{k}^{(\alpha)}.  
\end{equation}
By taking into account such constrained terms in the Hill-Wheeler equation
the stability of the solutions toward addition of eigenstates of 
the overlap kernel having small eigenvalues is found to be much increased,
and we can recognize the appearance of a wide range of plateaus
even in such a high angular momentum  region as $J=18\hbar$(Left part of Fig.1). 
The energy differences among the solutions are within the range of 
10keV$\sim$100keV.
This feature differs from our previous work[\ref{HOO95}].

\noindent
\begin{figure}[hhht]
\unitlength1mm
\begin{picture}(125,55)
\put(0,0){\epsfig{file=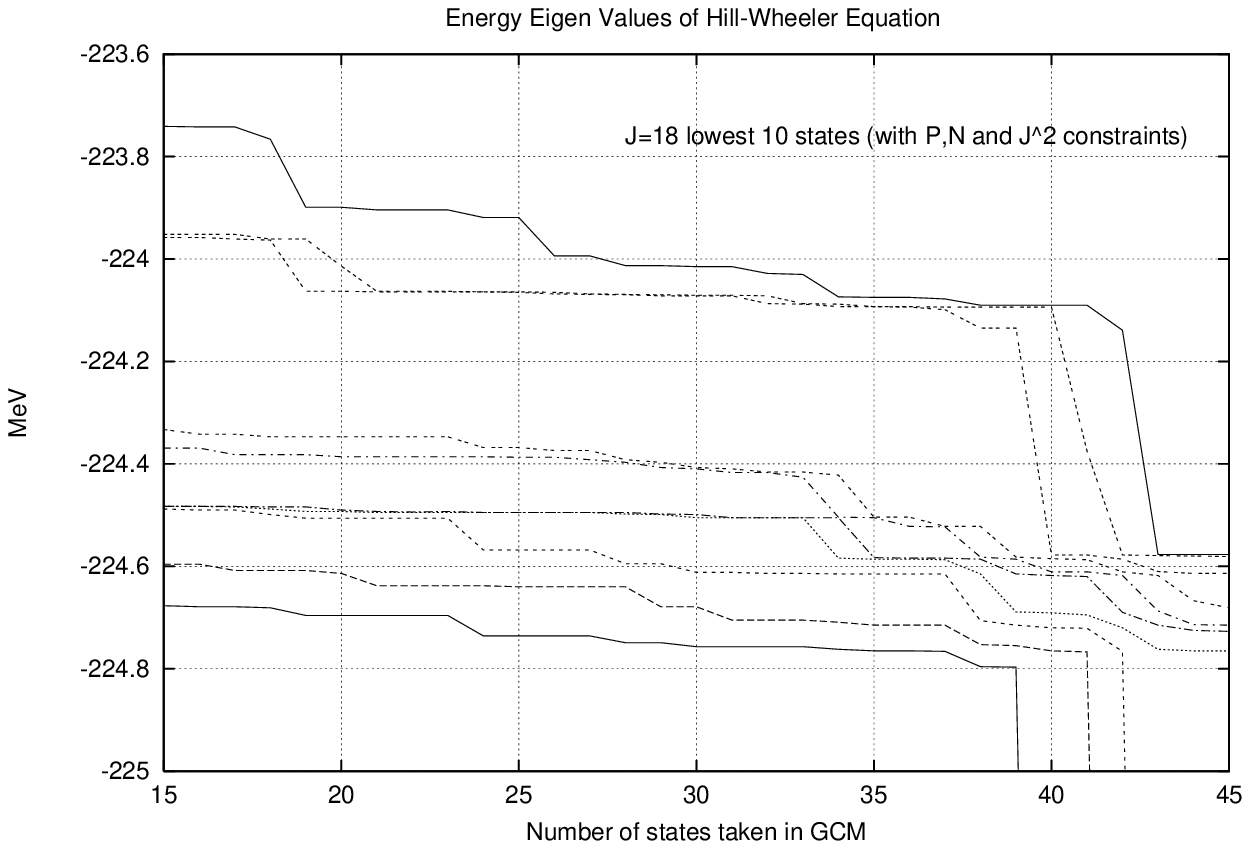,height=50mm}}
\put(80,0){\epsfig{file=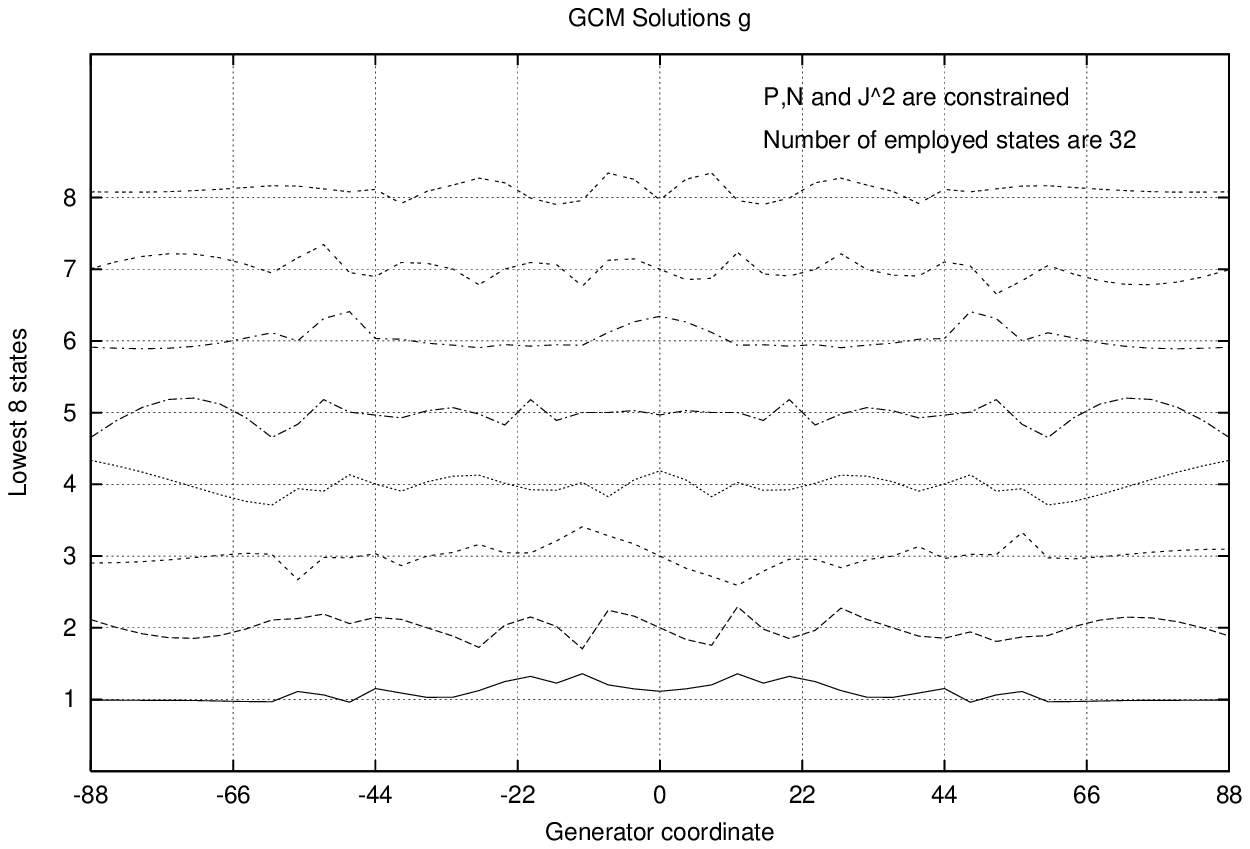,height=50mm}}
\end{picture}
\caption{Left part indicates the several lowest solutions of equation (1).
 Right part indicates the corresponding probability distributions of the
 GCM solutions.}
\end{figure} 

From the analysis of the wave function of the GCM solution,
we can observe several characteristic points. 
The behavior of the wave functions of each solution show an interesting
features (Right part of Fig.1).
One can see that the GCM wave functions are either symmetric or
anti-symmetric with respect to north and south latitudes.
This means that the broken symmetry of signature in TAR wave function is
recovered through the wobbling motion.
From the behavior of the lowest solution we can see that the $J=18\hbar$
state is a tilted-rotor in the present set of parameters. 
Furthermore, we noticed that the forth and the sixth solutions have an 
s$_{\rm n}$-band character.
 
This result is interesting since we have solved the GCM on
the TAR potential curve having minimum points located distant from the 
principal axis. 
If such a PAR solution will come into the lowest state,
that is an yrast, in the higher angular momentum region,
the situation will become much interesting because of the following reason.
In Fig.2, we show the existence of another 3D-CHFB 
solution along the prime meridian in this nucleus[\ref{HO97}]. 
Such solutions have been obtained by a calculation starting from the point 
$\theta=30^{\circ}$ back to the point $\theta=0^{\circ}$. 
Along this new solution the characters of proton and neutron gap
parameters  exchange the position from each other. 
Namely, $\Delta_p$ almost vanishes while $\Delta_n$ grows up to
a finite value at the PAR point.
This fact suggests a new type of seesaw vibrational mode of the
proton and neutron pairing, occurring through wobbling motion. 

In conclusion, we found several signatures of multi-band crossing
in this  nucleus.
The band mixing will become a {\em key} concept in understanding the 
band crossing mechanism of nuclei in this mass region.

\noindent
\begin{figure}[hh]
\unitlength1mm
\begin{picture}(125,55)
\put(35,0){\epsfig{file=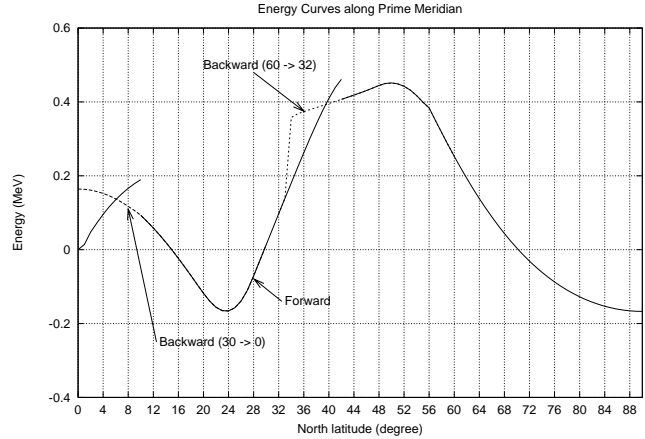,height=60mm}}
\end{picture}
\caption{The another 3D-CHFB solution (indicated ``Backward'' in the figure)
 along the prime meridian. Along this new solution the characters of proton and neutron gap parameters  exchange the position from each other.}
\end{figure} 


\noindent
{\bf References}
\begin{enumerate}
\setlength{\itemsep}{-0.1cm}
\setlength{\parsep}{0.0cm}
\item\label{Wa93}
P.M.Walker, Proc. Future of Nuclear Spectroscopy, Crete (1993).
\item\label{Sh97} 
T. Shizuma, {\it private communication}.
\item\label{HO94}
T. Horibata and N. Onishi,  Phys. Lett. B 325 (1994) 283, 
Nucl. Phys. A596 (1996) 251. 
\item\label{HOO95}
T. Horibata, M. Oi and N. Onishi, Phys. Lett. B355 (1995) 433.
\item\label{NLM76}
A. Neskakis et al., Nucl. Phys. A261 (1976) 198.
\item\label{BFH90}
P. Bonche et al., Nucl. Phys. A510 (1990) 466. 
\item\label{HO97}
T. Horibata and N. Onishi, RIKEN Accel. Prog. Rep. 31 (1997) 22.
\end{enumerate}
\end{document}